\documentclass[sn-mathphys,Numbered]{sn-jnl}

\usepackage{graphicx}%
\usepackage{multirow}%
\usepackage{amsmath,amssymb,amsfonts}%
\usepackage{amsthm}%
\usepackage{mathrsfs}%
\usepackage[title]{appendix}%
\usepackage{xcolor}%
\usepackage{textcomp}%
\usepackage{manyfoot}%
\usepackage{booktabs}%
\usepackage{algorithm}%
\usepackage{algorithmicx}%
\usepackage{algpseudocode}%
\usepackage{listings}%
\usepackage{mathtools,amsmath,amssymb,amsfonts,dsfont,mathrsfs,amsthm,latexsym}
\usepackage{graphicx}
\usepackage{xcolor}
\usepackage{siunitx}
\usepackage{braket}
\usepackage{xparse}
\usepackage{breqn}
\usepackage{tikz-cd}

\DeclareDocumentCommand{\Ag}{ s o }{ \IfBooleanTF{#1}
    { \IfValueTF{#2}{ \bm{\mathcal{A}}_{(#2)} }{ \bm{\mathcal{A}} } }
    { \IfValueTF{#2}{    {\mathcal{A}}_{(#2)} }{    {\mathcal{A}} } } }
\DeclareDocumentCommand{\Af}{ s o }{ \IfBooleanTF{#1}
    { \IfValueTF{#2}{ \boldsymbol{A}_{(#2)} }{ \boldsymbol{A} } }
    { \IfValueTF{#2}{            {A}_{(#2)} }{            {A} } } }
\DeclareDocumentCommand{\Fg}{ s o }{ \IfBooleanTF{#1}
    { \IfValueTF{#2}{ \bm{\mathcal{F}}_{(#2)} }{ \bm{\mathcal{F}} } }
    { \IfValueTF{#2}{    {\mathcal{F}}_{(#2)} }{    {\mathcal{F}} } } }
\DeclareDocumentCommand{\Ff}{ s o }{ \IfBooleanTF{#1}
    { \IfValueTF{#2}{ \boldsymbol{F}_{(#2)} }{ \boldsymbol{F} } }
    { \IfValueTF{#2}{            {F}_{(#2)} }{            {F} } } }

\newcommand{\ga}{\gamma}

\newcommand{\Mi}{\mathcal{M}}

\DeclareDocumentCommand{\PB}{ O{m} O{q} O{p} m m }{ \frac{ \partial #4 }{\partial {#2}^{#1} } \frac{ \partial #5 }{\partial {#3}_{#1} } - \frac{ \partial #4 }{\partial {#3}_{#1} } \frac{ \partial #5 }{\partial {#2}^{#1} } }

\DeclareDocumentCommand\Te{o o O{\,} m }{{T}_{#3}{}^{#1}_{#2}(#4)}

\NewDocumentCommand\MyAc{ m }{#1}
\DeclareDocumentCommand{\vif}{ t. t, t- s s m }{
  \RenewDocumentCommand\MyAc{ m }{##1}
  \IfBooleanT{#1}{\RenewDocumentCommand\MyAc{ m }{ \mathring{##1} } }
  \IfBooleanT{#2}{\RenewDocumentCommand\MyAc{ m }{ \tilde{##1} } }
  \IfBooleanT{#3}{\RenewDocumentCommand\MyAc{ m }{ \bar{##1} } }
  \IfBooleanTF{#4}
  { \IfBooleanTF{#5} { \hat{\MyAc{\boldsymbol{e}}}^{\hat{#6}} }{ \hat{\MyAc{\boldsymbol{e}}}^{{#6}} } }
  { \MyAc{\boldsymbol{e}}^{{#6}} } }
\DeclareDocumentCommand{\vi}{ t. t, t- s s m m}{
  \RenewDocumentCommand\MyAc{ m }{##1}
  \IfBooleanT{#1}{\RenewDocumentCommand\MyAc{ m }{ \mathring{##1} } }
  \IfBooleanT{#2}{\RenewDocumentCommand\MyAc{ m }{ \tilde{##1} } }
  \IfBooleanT{#3}{\RenewDocumentCommand\MyAc{ m }{ \bar{##1} } }
  \IfBooleanTF{#4}
  { \IfBooleanTF{#5} { \hat{\MyAc{e}}^{\hat{#6}}_{\hat{#7}} }{ \hat{\MyAc{e}}^{#6}_{{#7}} } }
  { \MyAc{e}^{{#6}}_{{#7}} } }

\DeclareDocumentCommand{\bt}{ t. t, t- s s m m m }{
  \RenewDocumentCommand\MyAc{ m }{##1}
  \IfBooleanT{#1}{\RenewDocumentCommand\MyAc{ m }{ \mathring{##1} } }
  \IfBooleanT{#2}{\RenewDocumentCommand\MyAc{ m }{ \tilde{##1} } }
  \IfBooleanT{#3}{\RenewDocumentCommand\MyAc{ m }{ \bar{##1} } }
  \IfBooleanTF{#4}
  { \IfBooleanTF{#5} { \hat{\MyAc{\mathcal{B}}}_{{#6}}{}^{\hat{#7}}{}_{\hat{#8}} }{ \hat{\MyAc{\mathcal{B}}}_{{#6}}{}^{{#7}}{}_{{#8}} } }
  { \MyAc{\mathcal{B}}_{{#6}}{}^{{#7}}{}_{{#8}} } }

\DeclareDocumentCommand{\ct}{ t. t, t- s s m m m }{
  \RenewDocumentCommand\MyAc{ m }{##1}
  \IfBooleanT{#1}{\RenewDocumentCommand\MyAc{ m }{ \mathring{##1} } }
  \IfBooleanT{#2}{\RenewDocumentCommand\MyAc{ m }{ \tilde{##1} } }
  \IfBooleanT{#3}{\RenewDocumentCommand\MyAc{ m }{ \bar{##1} } }
  \IfBooleanTF{#4}
  { \IfBooleanTF{#5} { \hat{\MyAc{\Gamma}}_{{#6}}{}^{\hat{#7}}{}_{\hat{#8}} }{ \hat{\MyAc{\Gamma}}_{{#6}}{}^{{#7}}{}_{{#8}} } }
  { \MyAc{\Gamma}_{{#6}}{}^{{#7}}{}_{{#8}} } }
\DeclareDocumentCommand{\spif}{ t. t, t- s s m m }{
  \RenewDocumentCommand\MyAc{ m }{##1}
  \IfBooleanT{#1}{\RenewDocumentCommand\MyAc{ m }{ \mathring{##1} } }
  \IfBooleanT{#2}{\RenewDocumentCommand\MyAc{ m }{ \tilde{##1} } }
  \IfBooleanT{#3}{\RenewDocumentCommand\MyAc{ m }{ \bar{##1} } }
  \IfBooleanTF{#4}
  { \IfBooleanTF{#5} { \hat{\MyAc{\boldsymbol{\omega}}}^{\hat{#6}}{}_{\hat{#7}} }{ \hat{\MyAc{\boldsymbol{\omega}}}^{{#6}}{}_{{#7}} } }
  { \MyAc{\boldsymbol{\omega}}^{{#6}}{}_{{#7}} } }
\DeclareDocumentCommand{\spi}{ t. t, t- s s m m m }{
  \RenewDocumentCommand\MyAc{ m }{##1}
  \IfBooleanT{#1}{\RenewDocumentCommand\MyAc{ m }{ \mathring{##1} } }
  \IfBooleanT{#2}{\RenewDocumentCommand\MyAc{ m }{ \tilde{##1} } }
  \IfBooleanT{#3}{\RenewDocumentCommand\MyAc{ m }{ \bar{##1} } }
  \IfBooleanTF{#4}
  { \IfBooleanTF{#5} { \hat{\MyAc{{\omega}}}_{\hat{#6}}{}^{\hat{#7}}{}_{\hat{#8}} }{ \hat{\MyAc{{\omega}}}_{{#6}}{}^{{#7}}{}_{{#8}} } }
  { \MyAc{{\omega}}_{{#6}}{}^{{#7}}{}_{{#8}} } }

\DeclareDocumentCommand{\rif}{ t. t, t- s s m m }{
  \RenewDocumentCommand\MyAc{ m }{##1}
  \IfBooleanT{#1}{\RenewDocumentCommand\MyAc{ m }{ \mathring{##1} } }
  \IfBooleanT{#2}{\RenewDocumentCommand\MyAc{ m }{ \tilde{##1} } }
  \IfBooleanT{#3}{\RenewDocumentCommand\MyAc{ m }{ \bar{##1} } }
  \IfBooleanTF{#4}
  { \IfBooleanTF{#5} { \hat{\MyAc{\bm{\mathcal{R}}}}{}^{\hat{#6}}{}_{\hat{#7}} }{ \hat{\MyAc{\bm{\mathcal{R}}}}{}^{{#6}}{}_{{#7}} } }
  { \MyAc{\bm{\mathcal{R}}}{}^{{#6}}{}_{{#7}} } }
\DeclareDocumentCommand{\ri}{ t. t, t- s s m m m }{
  \RenewDocumentCommand\MyAc{ m }{##1}
  \IfBooleanT{#1}{\RenewDocumentCommand\MyAc{ m }{ \mathring{##1} } }
  \IfBooleanT{#2}{\RenewDocumentCommand\MyAc{ m }{ \tilde{##1} } }
  \IfBooleanT{#3}{\RenewDocumentCommand\MyAc{ m }{ \bar{##1} } }
  \IfBooleanTF{#4}
  { \IfBooleanTF{#5} { \hat{\MyAc{\mathcal{R}}}_{{#6}}{}^{\hat{#7}}{}_{\hat{#8}} }{ \hat{\MyAc{\mathcal{R}}}_{{#6}}{}^{{#7}}{}_{{#8}} } }
  { \MyAc{\mathcal{R}}_{{#6}}{}^{{#7}}{}_{{#8}} } }

\DeclareDocumentCommand{\kf}{ t. t, t- s s m m }{
  \RenewDocumentCommand\MyAc{ m }{##1}
  \IfBooleanT{#1}{\RenewDocumentCommand\MyAc{ m }{ \mathring{##1} } }
  \IfBooleanT{#2}{\RenewDocumentCommand\MyAc{ m }{ \tilde{##1} } }
  \IfBooleanT{#3}{\RenewDocumentCommand\MyAc{ m }{ \bar{##1} } }
  \IfBooleanTF{#4}
  { \IfBooleanTF{#5} { \hat{\MyAc{\bm{\mathcal{K}}}}^{\hat{#6}}{}_{\hat{#7}} }{ \hat{\MyAc{\bm{\mathcal{K}}}}^{{#6}}{}_{{#7}} } }
  { \MyAc{\bm{\mathcal{K}}}^{{#6}}{}_{{#7}} } }
\DeclareDocumentCommand{\ko}{ t. t, t- s s m m m }{
  \RenewDocumentCommand\MyAc{ m }{##1}
  \IfBooleanT{#1}{\RenewDocumentCommand\MyAc{ m }{ \mathring{##1} } }
  \IfBooleanT{#2}{\RenewDocumentCommand\MyAc{ m }{ \tilde{##1} } }
  \IfBooleanT{#3}{\RenewDocumentCommand\MyAc{ m }{ \bar{##1} } }
  \IfBooleanTF{#4}
  { \IfBooleanTF{#5} { \hat{\MyAc{\mathcal{K}}}_{\hat{#6}}{}^{\hat{#7}}{}_{\hat{#8}} }{ \hat{\MyAc{\mathcal{K}}}_{{#6}}{}^{{#7}}{}_{{#8}} } }
  { \MyAc{\mathcal{K}}_{{#6}}{}^{{#7}}{}_{{#8}} } }

\DeclareDocumentCommand{\tf}{ t. t, t- s s m }{
  \RenewDocumentCommand\MyAc{ m }{##1}
  \IfBooleanT{#1}{\RenewDocumentCommand\MyAc{ m }{ \mathring{##1} } }
  \IfBooleanT{#2}{\RenewDocumentCommand\MyAc{ m }{ \tilde{##1} } }
  \IfBooleanT{#3}{\RenewDocumentCommand\MyAc{ m }{ \bar{##1} } }
  \IfBooleanTF{#4}
  { \IfBooleanTF{#5} { \hat{\MyAc{\bm{\mathcal{T}}}}^{\hat{#6}} }{ \hat{\MyAc{\bm{\mathcal{T}}}}^{{#6}} } }
  { \MyAc{\bm{\mathcal{T}}}^{{#6}} } }
\DeclareDocumentCommand{\tt}{ t. t, t- s s m m m }{
  \RenewDocumentCommand\MyAc{ m }{##1}
  \IfBooleanT{#1}{\RenewDocumentCommand\MyAc{ m }{ \mathring{##1} } }
  \IfBooleanT{#2}{\RenewDocumentCommand\MyAc{ m }{ \tilde{##1} } }
  \IfBooleanT{#3}{\RenewDocumentCommand\MyAc{ m }{ \bar{##1} } }
  \IfBooleanTF{#4}
  { \IfBooleanTF{#5} { \hat{\MyAc{\mathcal{T}}}_{\hat{#6}}{}^{\hat{#7}}{}_{\hat{#8}} }{ \hat{\MyAc{\mathcal{T}}}_{{#6}}{}^{{#7}}{}_{{#8}} } }
  { \MyAc{\mathcal{T}}_{{#6}}{}^{{#7}}{}_{{#8}} } }

\NewDocumentCommand{\MyLe}{}{}
\DeclareDocumentCommand{\PG}{ s s O{\Pi} m m m }{
  \RenewDocumentCommand\MyLe{}{\Gamma}
  \IfBooleanT{#1}{\RenewDocumentCommand{\MyLe}{}{ \bt{ }{ }{ } }}
  \IfBooleanT{#2}{\RenewDocumentCommand{\MyLe}{}{\Ag}}
  { {#3}_{\MyLe}{}^{#4}{}_{#5}{}^{#6} }
}

\newcommand*{\de}[1]{\mathop{\mathrm{d}#1}\nolimits}

\usepackage{hyperref}
\usepackage{wrapfig}
\usepackage{rotating}
\usepackage{breqn}

\begin{document}

\title[Article Title]{Does the metric play a fundamental role in the building of gravitational models?}


\author*[1,2]{\fnm{Oscar} \sur{Castillo-Felisola}}\email{o.castillo.felisola@protonmail.com}
\equalcont{These authors contributed equally to this work.}

\author*[1]{\fnm{Jos\'e} \sur{Perdiguero G\'arate}}\email{jose.perdiguerog@gmail.com}
\equalcont{These authors contributed equally to this work.}


\affil*[1]{\orgdiv{Departamento de F\'isica}, \orgname{Universidad T\'ecnica Federico Santa Mar\'ia}, \orgaddress{\street{Casilla 110-V}, \city{Valpara\'iso}, \postcode{2390123}, \state{Valpara\'iso}, \country{Chile}}}

\affil[2]{\orgname{Centro Cient\'ifico Tecnol\'ogico de Valpara\'iso}, \orgaddress{\street{Casilla 110-V}, \city{Valpara\'iso}, \postcode{2390136}, \state{Valpara\'iso}, \country{Chile}}}



\abstract{The idea that General Relativity could be an effective model, of a yet unknown theory of gravity, has gained momentum among theoretical
physicists. The \emph{polynomial affine model of gravity} is an alternative
model of affine gravity that possesses many desirable features to
pursue a quantum theory of gravitation. In this paper we argue that
such features are a consequence of the lack of a metric structure in
the building of the model, even though a \emph{emergent} metric could be
defined. The model introduces additional degrees of freedom associated
to the geometric properties of the space, which might shed light to
understand the nature of the dark sector of the Universe. When the
model is coupled to a scalar field, it is possible to define
inflationary scenarios.}

\keywords{Purely affine gravity, Cosmological model, Alternative gravitational models, Affine connection}



\maketitle

\section{Introduction}\label{sec1}

Nowadays the gravitational interaction is understood as the effect of
geometric properties of the spacetime. The conceptual break through
from the Newtonian interpretation of gravitational force to this
modern approach is due to Albert Einstein, who obtained the field
equations of General Relativity in an attempt to make the Newtonian
gravity compatible with the ideas of Special Relativity.

The Einstein field equations for gravity are expressed as
\begin{equation}
  \mathcal{R}_{\mu\nu} - \frac{1}{2} g_{\mu\nu} \mathcal{R} + \Lambda g_{\mu\nu} = 8 \pi G_N T_{\mu\nu},
  \label{eq:einstein-equations}
\end{equation}
where \(g_{\mu\nu}\) is the metric of the spacetime, \(G_N\) and
\(\Lambda\) are the Newton's gravitational and cosmological constants
respectively, the \(T_{\mu\nu}\) is the energy-momentum tensor of the
matter populating the Universe, and finally the \(\mathcal{R}\)
quantities are curvatures (which are calculable for a given metric).

David Hilbert obtained the field equations in Eq.
\eqref{eq:einstein-equations} through the optimisation of an action
functional, dubbed the Einstein--Hilbert action,
\begin{equation}
  S_{\text{EH}} = \int \mathrm{d}^4 x \, \sqrt{\vert g\vert} \left( \frac{1}{2 \kappa} \left( g^{\mu\nu} \mathcal{R}_{\mu\nu} - 2 \Lambda \right) + \mathcal{L}_m \right),
  \label{eq:einstein-hilbert-action}
\end{equation}
with \(\kappa = 8 \pi G_N\), \(g^{\mu\nu}\) is the \emph{inverse} of the
metric in the sense that \(g^{\mu\nu} g_{\nu\lambda} =
\delta^{\mu}_{\lambda}\), the symbol \(g\) denotes the determinant of
the metric tensor field, and finally the $\mathcal{L}_m$ refers to the
Lagrangian of the matter content, and it is related to the
energy-momentum tensor by the relation
\begin{equation}
  T_{\mu\nu} = - \frac{2}{\sqrt{|g|}} \frac{\partial}{\partial g^{\mu\nu}} \sqrt{|g|} \mathcal{L}_m.
  \label{eq:energy-momentum-def}
\end{equation}

The above formulation of gravity relies on the premise that the
spacetime is modelled by a pseudo-Riemannian manifold whose metric is
Lorentzian with signature \(2\). Consequently, the curvatures involved
in Eqs. \eqref{eq:einstein-equations} and
\eqref{eq:einstein-hilbert-action} are obtained from the Riemannian
curvature defined by the Levi-Civita connection, which is completely
determined by the metric (the explicit formulas can be found in every
textbook on General Relativity).

Despite its success explaining a large amount of observations
\cite{will14_confr_between_gener_relat,will18_theor_exper_gravit_physic,abbott16_obser_gravit_waves_from_binar,abbott16_gw151},
from a theoretical perspective, two of the major unappealing points of
General Relativity are that it fails to be consistently quantised and
that---in order to match the cosmological observations---requires a
large amount of (about 96\% of the total matter/energy of the Universe)
matter/energy whose nature is almost completely unknown, grouped under
the denomination of \emph{dark sector}. These issues serve as motivation to
explore alternative models of gravity.

A customary way to propose generalisations to the Einstein--Hilbert
formulation of gravity is to follow the assertion of the Lagrangian
formulation of classical mechanics (classical field theory, and in
particular effective theories), ``All terms allowed by the symmetries
of your system should be included in the action functional''. This
method has lead us to consider models whose Lagrangians include higher
powers in curvature, such as Lanczos--Lovelock 
\cite{lanczos38_remar_proper_rieman_chris_tensor_four,lovelock69_uniquen_einst_field_equat_four_dimen_space,lovelock71_einst_tensor_its_gener},
or the \(f(\mathcal{R})\) \cite{de10_r,sotiriou10_r_theor_of_gravit},
and other extensions
\cite{salvio18_quadr_gravit,oliva10_new_cubic_theor_gravit_five_dimen,bueno16_einst_cubic_gravit}.

There are other type of extensions, guided by the proposal by
Palatini, in which the metric and the connection are considered as
independent fields \cite{palatini19_deduz_invar_delle_equaz_gravit}.

Formally speaking, the approach \emph{a la Palatini} is supported by the fact
that on a manifold the affine and metric structures are not related in
general. We could use an analogy with the classical plane geometry,
where the compass and ruler are the tools defining the geometry.
Continuing with the analogy, the compass yields the notion of distance
(i.e. it is the equivalent to the metric in differential geometry),
while the ruler is used to define parallelism (i.e. it is the
analogous to the affine connection in differential geometry). Despite
the fact that we could use a graduated ruler to build a geometry with
a single instrument---this would be the analogous to Riemannian
geometry---, it is possible to relax the conditions and consider
different pairs of instruments, including the possibility of using
only one of them (such are the cases of \emph{inversive} and \emph{projective}
geometry).

A manifold endowed with an affine connection,
\((\mathcal{M},\hat{\Gamma})\), is called an \emph{affinely connected
manifold}. In addition, if a metric structure is allowed on the
manifold, \((\mathcal{M},\hat{\Gamma},g)\), it is said to be
metric-affine. Any pair \((\hat{\Gamma},g)\) is characterised by the
curvature of the connection, \(\mathcal{R}\), whether the connection
possesses an skew-symmetric part (also known as torsion tensor),
\(\mathcal{T}\), and the failure of compatibility
\(\nabla^{\hat{\Gamma}} g = \mathcal{Q}\). There are eight types of
metric-affine manifolds, e.g. Weyl, Weyl--Cartan, Weitzenböck,
Riemann--Cartan, Riemann, Minkowski and others (see Refs.
\cite{hehl95_metric_affin_gauge_theor_gravit,castillo-felisola21_aspec_polyn}).
The commutative diagram in Eq. \eqref{eq:type-of-manifolds} illustrate
the diversity of metric-affine manifold, which can be obtained from
the class of affinely connected ones.

\begin{equation}
  \begin{tikzcd}[row sep=scriptsize, column sep=scriptsize]
    & (\Mi,\hat{\Gamma}) \ar[d, "g"] & & \\
    & (\mathcal{Q},\mathcal{R},\mathcal{T}) \arrow[dl] \arrow[rr] \arrow[dd] & & (\mathcal{Q},\mathcal{T}) \arrow[dl] \arrow[dd] \\
    (\mathcal{Q},\mathcal{R}) \arrow[rr, crossing over] \arrow[dd] & & (\mathcal{Q}) \\
    & (\mathcal{R},\mathcal{T}) \arrow[dl] \arrow[rr] & & (\mathcal{T}) \arrow[dl] \\
    (\mathcal{R}) \arrow[rr] & & (\text{Flat}) \arrow[from=uu, crossing over]\\
  \end{tikzcd}
  \label{eq:type-of-manifolds}
\end{equation}

In the literature, one can find a plethora of proposals in which the
spacetime is modelled by any of the kind of metric-affine manifolds.
Just to mention a few of them, Weyl considered a spacetime with
vectorial non-metricity
\cite{weyl18_gravit_und_elekt,weyl18_reine_infin,weyl21_zur_infin,weyl22_space},
Cartan considered a connection with nonvanishing torsion
\cite{cartan22_sur_une_de_la_notion,cartan23_sur_les_connex_affin_et,cartan24_sur_les_connex_affin_et,cartan25_sur_les_connex_affin_et},
Einstein proposed a version of General Relativity with a flat
connection with torsion (a Weitzenböck connection)
\cite{simon06_alber_einst,aldrovandi13_telep_gravit,maluf13_telep_equiv_gener},
extension of the Lanczos--Lovelock to Riemann--Cartan manifolds
\cite{mardones91_lovel}, models with higher powers in curvature,
torsion and/or non-metricity
\cite{capozziello10_metric_affin_r_with_torsion,capozziello11_exten_theor_gravit,harko11_r_t,katirci14_r_t_gravit_cardas_like,cai16_t_telep_gravit_cosmol,jimenez18_telep_palat_theor,bombacigno18_r_gravit_with_torsion_immir_field,wang18_ricci_gravit,iosifidis19_metric_affin_gravit_cosmol_torsion,jimenez-cano20_new_metric_affin_gener_gravit_wave_geomet,xu20_weyl_type_q_t_gravit,arora20_const_q_t_gravit_from_energ_condit,jimenez20_cosmol_q_geomet,jimenez-cano22_metric_affin_gauge_theor_gravit,jimenez22_parit_odd_sector_metric}.

Affine formulations of gravity, i.e. models built up on
\((\mathcal{M},\hat{\Gamma})\),\footnote{Those do not require the existence of a fundamental metric and
whose mediating field are affine connections.} bringing the gravitational
interaction to the \emph{same footage} as gauge theories---since their
fundamental field is a connection---,\footnote{The gauge symmetry is a key ingredient in the quantisation and
renormalisation properties of the other fundamental interactions.} and therefore are
believed to play a major role in the development of a quantum theory
of gravity. Moreover, the mismatch in the number of independent
components (metric to connection) could generate novel gravitational
degrees of freedom, which might provide \emph{geometrical} candidates of dark
matter/energy.\footnote{In a \(D\)-dimensional space(time), the number of components
of the affine connection is \(D^3\), which supersize the number of
components of a standard metric (\(\frac{D(D+1)}{2}\)) or even
\emph{generalised} metrics (\(D^2\)).} Hence, from a theoretical point of view, the
analysis of affine models of gravity is very interesting.
Historically, the first affine models were built by Einstein,
Eddington and Schrödinger
\cite{einstein23_zur_affin_feldt,einstein23_theor_affin_field,eddington23,schroedinger50_space},
but more recent models were considered by Kijowski
\cite{kijowski78_new_variat_princ_gener_relat,kijowski07_univer_affin_formul_gener_relat},
Popławski
\cite{poplawski07_nonsy_purel_affin,poplawski07_unified_purel_affin_theor_gravit_elect,poplawski14_affin_theor_gravit},
Krasnov
\cite{krasnov07_non_metric_gravit,krasnov11_pure_connec_action_princ_gener_relat,delfino15_pure_connec_formal_gravit_lin,delfino15_pure_connec_formal_gravit_feyn},
and ourselves
\cite{castillo-felisola15_polyn_model_purel_affin_gravit,castillo-felisola18_einst_gravit_from_polyn_affin_model}.

\section{Formulation of Polynomial Affine Gravity}\label{sec2}

The polynomial affine model of gravity is built up with powers of the
affine connection, while the role of the metric field is dismisses by
keeping the action independent of it (similar to a Schwarz type
topological theory \cite{birmingham91_topol_field_theor})
\cite{castillo-felisola15_polyn_model_purel_affin_gravit,castillo-felisola18_einst_gravit_from_polyn_affin_model}.

The action for this model is the most general functional built up with
the \emph{irreducible} components of the affine connection and a volume
form. Mathematically, a general affine connection
\(\ct*{\mu}{\lambda}{\nu}\) decomposes as follows,\footnote{Without evoking to a metric, it is not possible to separate the
Levi-Civita and non-metricity contributions from the symmetric
connection.} 
\begin{equation}
  \ct*{\mu}{\lambda}{\nu} 
  = \ct*{(\mu}{\lambda}{\nu)} + \ct*{[\mu}{\lambda}{\nu]}
  = \ct{\mu}{\lambda}{\nu} + \bt{\mu}{\lambda}{\nu} + \Ag_{[\mu}\delta^\lambda_{\nu]},
  \label{eq:conn_decomp}
\end{equation}
where \(\ct{\mu}{\lambda}{\nu}\) denotes the symmetric part of the
connection, and \(\bt{\mu}{\lambda}{\nu}\) and \(\Ag_\mu\) associated
with the traceless and trace components of the torsion.\footnote{The symmetric connection might be split further into
irreducible components, but such separation is not necessary for the
subsequent analysis.} Since
the symmetric connection is the sole \emph{irreducible} piece that is not a
tensor, it enters into the action functional through the covariant
derivative or Chern--Simons topological terms. Furthermore, we shall
consider a \emph{canonical} volume form (i.e. an everywhere nonvanishing
\(m\)-form on an \(m\)-dimensional affine manifold) defined by,
\(\de{V}^{\mu_1 \cdots \mu_n} = \de{x}^{\mu_1} \wedge \cdots \wedge
\de{x}^{\mu_m}\).

Hence, with the aids of an \emph{index structural analysis},\footnote{For details on this analysis see Refs.
\cite{castillo-felisola18_einst_gravit_from_polyn_affin_model}.} the most
general actions have been built up in three and four dimensions (see
Refs.
\cite{castillo-felisola15_polyn_model_purel_affin_gravit,castillo-felisola18_einst_gravit_from_polyn_affin_model,castillo-felisola18_cosmol,castillo-felisola20_emerg_metric_geodes_analy_cosmol,castillo-felisola_pag3d}),
and up to topological and boundary terms.\footnote{In the three-dimensional action we have included the
Chern--Simons term, which has played a fundamental role in the
development of three-dimensional gravity and its possible quantum
programme
\cite{witten88_dimen_gravit_as_exact_solub_system,horne89_confor_gravit_three_dimen_as_gauge_theor,witten07_three_dimen_gravit_revis,carlip98_quant,mielke91_topol_gauge_model_gravit_with_torsion,mielke07_s_dualit_gravit_with_torsion}.} In the following we
shall restrict ourselves to the four-dimensional case, whose action is
\begin{dmath}
  \label{eq:action_4d}
  S_{(4)}  = \int \de{V}^{\alpha \beta \gamma \delta} \bigg[
  B_1 \ri{\mu\nu}{\mu}{\rho} \bt{\alpha}{\nu}{\beta} \bt{\gamma}{\rho}{\delta}
  + B_2 \ri{\alpha\beta}{\mu}{\rho} \bt{\gamma}{\nu}{\delta} \bt{\mu}{\rho}{\nu}
  + B_3 \ri{\mu\nu}{\mu}{\alpha} \bt{\beta}{\nu}{\gamma} \Ag_\delta 
  + B_4 \ri{\alpha\beta}{\sigma}{\rho} \bt{\gamma}{\rho}{\delta} \Ag_\sigma
  + B_5 \ri{\alpha\beta}{\rho}{\rho} \bt{\gamma}{\sigma}{\delta} \Ag_\sigma 
  + C_1 \ri{\mu\alpha}{\mu}{\nu} \nabla_\beta \bt{\gamma}{\nu}{\delta} 
  + C_2 \ri{\alpha\beta}{\rho}{\rho} \nabla_\sigma \bt{\gamma}{\sigma}{\delta} 
  + D_1 \bt{\nu}{\mu}{\lambda} \bt{\mu}{\nu}{\alpha} \nabla_\beta \bt{\gamma}{\lambda}{\delta}
  + D_2 \bt{\alpha}{\mu}{\beta} \bt{\mu}{\lambda}{\nu} \nabla_\lambda \bt{\gamma}{\nu}{\delta} 
  + D_3 \bt{\alpha}{\mu}{\nu} \bt{\beta}{\lambda}{\gamma} \nabla_\lambda \bt{\mu}{\nu}{\delta} 
  + D_4 \bt{\alpha}{\lambda}{\beta} \bt{\gamma}{\sigma}{\delta} \nabla_\lambda \Ag_\sigma 
  + D_5 \bt{\alpha}{\lambda}{\beta} \Ag_\sigma \nabla_\lambda \bt{\gamma}{\sigma}{\delta}
  + D_6 \bt{\alpha}{\lambda}{\beta} \Ag_\ga \nabla_\lambda \Ag_\delta 
  + D_7 \bt{\alpha}{\lambda}{\beta} \Ag_\lambda \nabla_\ga \Ag_\delta 
  + E_1 \nabla_\rho \bt{\alpha}{\rho}{\beta} \nabla_\sigma \bt{\gamma}{\sigma}{\delta} 
  + E_2 \nabla_\rho \bt{\alpha}{\rho}{\beta} \nabla_\ga \Ag_{\delta}
  + F_1 \bt{\alpha}{\mu}{\beta} \bt{\gamma}{\sigma}{\delta} \bt{\mu}{\lambda}{\rho} \bt{\sigma}{\rho}{\lambda} 
  + F_2 \bt{\alpha}{\mu}{\beta} \bt{\gamma}{\nu}{\lambda} \bt{\delta}{\lambda}{\rho} \bt{\mu}{\rho}{\nu} 
  + F_3 \bt{\nu}{\mu}{\lambda} \bt{\mu}{\nu}{\alpha} \bt{\beta}{\lambda}{\gamma} \Ag_\delta 
  + F_4 \bt{\alpha}{\mu}{\beta} \bt{\gamma}{\nu}{\delta} \Ag_\mu \Ag_\nu
  \bigg],
\end{dmath}
where the covariant derivation and the curvature are defined with
respect to the symmetric connection, i.e. \({\nabla = \nabla^\Gamma}\) 
and \(\ri{}{}{}=\ri{}{\Gamma}{}\).

These actions have many desirable features: (i) it is invariant under
the complete group of diffeomoephisms; (ii) it is power-counting
renormalisable, which does not assure the renormalisability of the
model, but hints to the right direction; (iii) since the action
contains all the possible terms compatible with the desired structure,
it possesses a quality we call \emph{rigidity}, due to the fact that (in
principle) any counter-terms should have the form of a term in the
action; (iv) all coupling constants are dimensionless, suggesting
conformal invariance of the model, at least at three level; (v)
possess sectors compatible with General Relativity, i.e. the field
equations are generalisations to those of General Relativity
\cite{castillo-felisola18_einst_gravit_from_polyn_affin_model,castillo-felisola20_emerg_metric_geodes_analy_cosmol};
(vi) the field equations derived from the action in Eq.
\eqref{eq:action_4d} are systems of non-linear, coupled, second order
partial differential equations for the components of the affine
connection \cite{castillo-felisola20_emerg_metric_geodes_analy_cosmol};
(vii) since the action depends on the first derivatives of the fields,
in principle there is no need of considering affine analogous to the
Gibbons–Hawking–York term;\footnote{In other models of affine gravity these additional terms are
required \cite{parattu16_bound_term_gravit_action_with_null_bound,krishnan17_robin_gravit,krishnan17_neuman_bound_term_gravit,lehner16_gravit_action_with_null_bound,hopfmueller17_gravit_degrees_freed_null_surfac,jubb17_bound_corner_terms_action_gener_relat}.} (vii) there are (to our knowledge)
three possible symmetric \(\binom{0}{2}\)-tensors derived from the
connection, which might be define metric structures on the manifold
\(\mathcal{M}\) if at least one of them is nondegenerated
\cite{castillo-felisola20_emerg_metric_geodes_analy_cosmol,castillo-felisola21_aspec_polyn_affin_model_gravit_three,castillo-felisola22_polyn_affin_model_gravit};
(viii) there is a \emph{natural} way to couple scalar fields to the
polynomial affine gravity, which under certain assumptions is shown to
be equivalent the minimally coupled Einstein--Klein--Gordon system
\cite{castillo-felisola18_einst_gravit_from_polyn_affin_model}.

Before proceeding, let us resume what we have obtained. The polynomial
affine model of gravity is a model whose fundamental field is the
affine connection, the action does not require the existence of a
fundamental metric tensor, however, it is possible to endow the
manifold with metric structures which are \emph{descendent} (also dubbed
\emph{derived}) from the connection.

For the sake of simplicity, we might focus on the torsion-free sector
of the model, i.e. with \(\mathcal{A} \to 0\) and \(\mathcal{B} \to 0\).
It has been shown that such sector is a consistent truncation of the model,
even though the Einstein-like equations in polynomial affine
gravity arise from the field equations of the \(\bt{}{}{}\)-field
(behaviour which has been attributed as a sign of the non-uniqueness
of the Lagrangian description of the system \cite{hojman_privat}, see
also a short discussion on Sec. 3 of Ref.
\cite{castillo-felisola18_einst_gravit_from_polyn_affin_model}). In
four dimensions, the effective field equations in the torsion-less
sector are
\begin{equation}
  \nabla_{[\lambda} \ri{\mu]\nu}{}{} = 0,
  \text{ or equivalently }
  \nabla_{\lambda} \ri{\mu\nu}{\lambda}{\rho} = 0.
  \label{eq:ricci-codazzi}
\end{equation}
The first expression in Eq. \eqref{eq:ricci-codazzi} is a Codazzi-like condition, and
therefore the Ricci tensor is said to be a Codazzi tensor
\cite{derdzinski83_codaz_tensor_field_curvat_pontr_forms,besse07_einst}.
The second expression in Eq. \eqref{eq:ricci-codazzi} is obtained from the former
after applying the differential Bianchi identity, and the curvature is
said to be harmonic
\cite{derdzinski80_class_certain_compac_rieman_manif,derdzinski82_compac_rieman_manif_with_harmon_curvat,derdzinski85_rieman}.

In addition, the equations in \eqref{eq:ricci-codazzi} are well-known
generalisations to the Einstein field equations, which appear as part
of the dynamics of a model of Stephenson--Kilmister--Yang model of gravity
\cite{stephenson58_quadr_lagran_gener_relat,kilmister61_use_alg_struct_phys,yang74_integ_formal_gauge_field},
which is a Yang--Mils formulation of gravity. Although the original
formulation of the Stephenson--Kilmister--Yang model have structural
incompatibilities with General Relativity, such issue is introduced by
the field equation for the metric field, and hence from the point of
view of polynomial affine gravity the problem is bypasses.

The cosmological solutions to the Eq. \eqref{eq:ricci-codazzi} are
obtained by proposing an ansatz for the connection
\cite{castillo-felisola18_beyond_einstein,castillo-felisola18_cosmol,castillo-felisola20_emerg_metric_geodes_analy_cosmol},
and the solutions might be classified in three types: (R.i) solutions
whose Ricci tensor vanishes; (R.ii) solutions whose Ricci tensor is
nontrivial and parallel; and (R.iii) solutions whose Ricci tensor is
nontrivial, non-parallel but it is a Codazzi tensor. Since the
symmetric part of the Ricci tensor (when no-degenerated) can be
interpreted as a metric tensor field \cite{schouten13_ricci},\footnote{It is worth mentioning that when the torsion is nonvanishing,
there is another tensor (derived from the connection) which can play
the role of a metric. We refer to such geometrical object as Popławski
metric tensor, in honour to the person who discover its relevance
\cite{poplawski14_affin_theor_gravit}.} it is
clear that the cosmological solutions  of type (R.ii) are equivalent to
Einstein manifolds, i.e. solutions to the Einstein field equations in
the vacuum, while the solutions of type (R.iii) extend the gravitational
sector of General Relativity by allowing non-metricity.

The consequences of the existence of such \emph{emergent} metric are vast.
Mainly, it introduce a notion of distance in the model, bringing it
closer to the physical world, on which we are able to give a \emph{global}
notion of distances between events.\footnote{In affinely connected spaces without metric, it is possible to
define distances along self-parallel curves, however, it is not
possible to compare the distances measured on different self-parallel
curves (see for example Ref. \cite{schroedinger50_space}). This quality
validate the use of the term \emph{global} in the previous sentence.} In addition, the metric
structure provides a notion of norm of vectors. In particular, if we
consider the norm of vectors which generate self-parallel curves,
their norm will differentiate the orbits corresponding to massless or
massive test particles.

Intriguingly, the signature of the emergent metrics is not fixed,
permitting the possibility of changes in the signature of the metric
(in cosmological scenarios as a function of time). Even though these
jumps in the signature require the degeneracy of the \emph{metric} on some
hyper-surface of the manifold, it remains to be analyses if such kind
of degeneracy is spurious (due to the choice of coordinate system for
example). However, such flexibility of the emergent metric play a
fundamental role if one would like to analyse the structure of the
saddle point of the path integral of gravity, in the spirit of
Witten's proposal \cite{witten21_note_compl}.

\section{Coupling a scalar field}\label{sec3}

Furthermore, in four dimensions the polynomial affine model of gravity
can be coupled to a scalar field \(\phi\)
\cite{castillo-felisola18_einst_gravit_from_polyn_affin_model}, through
the geometrical object defined as
\begin{dmath}
  \mathrm{g}^{\mu\nu} = \left(\alpha \nabla_\lambda \mathcal{B}_{\rho}{ }^{\mu}{ }_{\sigma} + \beta \mathcal{A}_\lambda 
  \mathcal{B}_{\rho}{ }^{\mu}{ }_{\sigma}\right)\mathrm{d}V^{\nu\lambda\rho\sigma} + \gamma \mathcal{B}_{\kappa}{ }^{\mu}{ }_{\lambda}
  \mathcal{B}_{\rho}{ }^{\nu}{ }_{\sigma}\mathrm{d}V^{\kappa\lambda\rho\sigma}.
  \label{eq:g-inverse}
\end{dmath}
Using the above expression we can define the kinetic term 
\begin{equation}
  S_{\phi} = - \int \mathrm{g}^{\mu\nu} \partial_{\mu} \phi \partial_{\nu} \phi,
  \label{eq:action-phi}
\end{equation}
which contribute to the field equations of the model even in the
torsion-free sector.

Furthermore, we can introduce a non-minimal coupling by the re-scaling
of the volume form,\footnote{This scaling is allowed since any non-degenerated \(4\)-form
is a valid volume form. In addition, the scaling by a function of the
scalar field, \(\phi\), could be interpreted as an analogous to the
deformation of the metric that allows to define non-linear sigma
models, i.e. \(g_{\mu\nu} \to g_{\mu\nu}(\phi)\).}
\begin{equation}
  \de{V}^{\mu\nu\lambda\rho} \mapsto \frac{ \de{V}^{\mu\nu\lambda\rho} }{\mathcal{V}(\phi)}.
  \label{eq:volume-scaling}
\end{equation}
The \(\mathcal{V}(\phi)\) function could be related to a
\emph{self-interacting} scalar field potential, turning this framework into a
natural ground for inflationary scenarios
\cite{kijowski07_univer_affin_formul_gener_relat,azri17_affin_inflat,azri18_induc_affin_inflat}.

The effective field equations in the torsion-free sector coupled with
a scalar field its given by
\begin{equation}
  \nabla_{[\lambda} \mathcal{S}_{\mu]\nu}  = 0,
\end{equation}
where the \(\mathcal{S}\) tensor its a symmetric tensor defined by
\begin{equation}
  \mathcal{S}_{\mu\nu} = \frac{\mathcal{R}_{\mu\nu} - C \partial_{\mu} \phi \partial_\nu \phi }{\mathcal{V}(\phi)},
\end{equation}
where the constant \(C\) measures the ratio between the original
coupling constants.

Just like the we mention before, we can classify solutions in three
different categories: (S.i) solutions whose \(\mathcal{S}\) tensor
vanishes; (S.ii) solutions where the \(\mathcal{S}\) tensor its
covariantly constant; and (S.iii) solutions where the \(\mathcal{S}\)
tensor its a codazzi tensor.

As a remark, when \(\mathcal{S}\) tensor its non-degenerate it endows
the manifold with an emergent metric tensor, however, it depends on
the scalar field, \(\phi\), similar as in the case of non-linear sigma
models. Note that if the Ricci tensor degenerates in a \emph{spatial}
component, such degeneracy would be inherited by the \(\mathcal{S}\)
tensor.

\section{Conclusion}\label{sec13}

Summarising, the polynomial affine gravity seems to be a viable model
of gravitational interactions, even if its formulation does not
require a metric structure. However, it is possible to induce a metric
structure using the affine connection, allowing a \emph{fair} way to compare
with the cosmological observations. Moreover, the lack of a functional
metric in the foundation of the model endows it with desirable
features, such as the \emph{rigidity}. 

We would like to emphasise that other
alternative models of gravity do not have restrictions to the number
of terms to be added to the action, and the arguments use to justify
why some terms should be ignored are bound only by the criteria of the
researchers, instead of driven by the theory itself. Also, the dark
sector of the Universe could be \emph{naturally} explained as by extra
degrees of freedom, which lie in the non-Riemannian part of the
connection. The cosmological constant in polynomial affine gravity is
introduced as an integration constant, changing the paradigm of its
interpretation similar to what happens in unimodular gravity and other
affine models (e.g. Einstein--Eddington--Schrödinger). 

In addition,
the laxity of the emergent metric might be important, in the light of
the proposal by Witten \cite{witten21_note_compl}, in a future
programme to quantise the model. Finally, given that there is a
\emph{natural} way to couple scalar fields (and enrich them with a
potential), it turns interesting to investigate inflationary scenarios
within the context of polynomial affine gravity. A couple of
interesting cases to consider are the power-law potential---for its
simplicity, and power structure like the gravitational sector---, and
the Starobinsky potential---given its success.

\backmatter





\bmhead{Acknowledgments}

The work OCF is sponsored by the “Centro Científico y Tecnológico de Valparaíso” (CCTVal), funded by the Chilean Government through the Centers of Excellence Base Financing Program of Agencia Nacional de Investigación y Desarrollo (ANID), by grant ANID PIA/APOYO AFB220004.

\bibliography{sn-bibliography}

\end{document}